\documentclass[12pt]{iopart}
\usepackage[numbers]{natbib}

\usepackage[dvips]{graphicx}
\usepackage[dvips,usenames]{color}

\newcommand{\avk}{\langle k\rangle}

\begin{document}

\title{Complex networks and glassy dynamics: walks in the energy landscape}

\author{Paolo Moretti$^1$, Andrea Baronchelli$^1$, Alain
  Barrat$^{2,3}$, and Romualdo Pastor-Satorras$^1$}
 
\address{$^1$ Departament de F\'\i sica i Enginyeria Nuclear,
  Universitat Polit\`ecnica de Catalunya, Campus Nord B4, 08034
  Barcelona, Spain}

\address{$^2$ Centre de Physique Th\'eorique (CNRS UMR 6207), Luminy,
  13288 Marseille Cedex 9, France} 

\address{$^3$ Complex Networks Lagrange Laboratory, Institute for
  Scientific Interchange (ISI), Torino, Italy}

\begin{abstract}
We present a simple mathematical framework for the description of the
dynamics of glassy systems in terms of a random walk in a complex
energy landscape pictured as a network of minima. We show how to use
the tools developed for the study of dynamical processes on complex
networks, in order to go beyond mean-field models that consider that
all minima are connected to each other. We consider several
possibilities for the transition rates between minima, and show that
in all cases the existence of a glassy phase depends on a delicate
interplay between the network's topology and the relationship between
energy and degree of a minimum. Interestingly, the network's degree
correlations and the details of the transition rates do not play any
role in the existence (nor in the value) of the transition
temperature, but have an impact only on more involved properties.  For
Glauber or Metropolis rates in particular, we find that the
low-temperature phase can be further divided into two regions with
different scaling properties of the average trapping time.  Overall,
our results rationalize and link the empirical findings about
correlations between the energy of the minima and their degree, and
should stimulate further investigations on this issue.
\end{abstract}


\maketitle

\section{Introduction}

In the last decade, studies about the structure and dynamics of
complex networks have blossomed, thanks in particular to the
versatility of the network representation, which has turned out to be
adequate for systems as diverse as the Internet or social networks. A
large body of knowledge about the empirical description of networked
systems has thus been accumulated, together with a wealth of modeling
techniques; a good level of understanding of how dynamical processes
taking place on networks depend on their structure has been as well
reached
\cite{Albert:2002,Dorogovtsev:2003,Newman:2003,Pastor:2004,Caldarelli:2007,Barrat:2008}.
Many network studies have been concerned with systems of interest in
several scientific areas a priori remote from physics (social
sciences, biology, computer science, epidemiology, ...), and they have
also reached more traditional fields of statistical physics, such as
the study of glassy systems, as we now describe.

The many puzzles raised by the glass transition, and in particular
the slow dynamics displayed by glassy systems
at low temperatures, have been the subject of a large interest
in the past decades \cite{Debenedetti_1,Debenedetti_2}. One of the
approaches which has led to promising insights
consists in the description of the dynamics of a glassy
system inside its configuration space. The energy landscape of a
glassy system is typically rugged, made of many local minima
(metastable states), whose huge number makes it difficult to reach
equilibrium. In this framework, the energy landscape is seen as a set
of basins of attractions of local minima (``traps''), and the system
evolves through a succession of harmonic vibrations inside traps 
and jumps between
minima \cite{Angelani:1998_1,Angelani:1998_2}. This picture has
stimulated the definition and study of various simplified models of
dynamical evolution between traps, in order to reproduce the
phenomenology of glassy dynamics
\cite{Bouchaud:1992,Bouchaud:1995,Barrat:1995,Monthus:1996,Bertin1,Bertin2,Bertin3}.
On the other hand, several studies have focused on obtaining a better
understanding of the structure of these local minima. A way to attain this goal is to perform
numerical simulations of small systems, at a fixed temperature, and
quenching them at regular time intervals in order to make them reach
the nearest local minimum. Information is then gathered on the various
local minima, and on the sizes of their basins of attraction.  Various
studies have investigated, among other issues, the detailed structure
of the potential energy landscape, the substructure of minima, and the
properties of energy barriers between minima
\cite{Heuer_1,Heuer_2,Heuer_3}. Several works have also used the
information on the energy landscape to study a master equation for the
time evolution of the probability to be in each minimum. The
considered systems range from clusters of Lennard-Jones atoms to
proteins or heteropolymers
\cite{Angelani:1998_1,Angelani:1998_2,Cieplak:1998_1,Cieplak:1998_2,Carmi:2009}.

An interesting property of the modeling of the energy landscape in
terms of a set of traps linked by energy barriers, lies in the
possibility to define and study its network representation within the
context of network theory. In this representation, each local minimum
is associated to a node, and a link is drawn between two nodes
whenever it is possible for the system to jump between the basins of
attraction of the corresponding minima. The links can then be defined
as weighted and directed, as jumps between minima are not
equiprobable, and may be easier in one direction than in the
other. Networks of local minima of the energy landscape have thus been
built and studied. These networks have been found to exhibit a
small-world character \cite{Scala:2001}. The number of links of each
node (its degree) turns out to be strongly heterogeneous, possibly
with scale-free degree distributions, which have been linked to
scale-free distributions of the areas of the basins of attraction
\cite{Doye:2002,Massen:2005,Seyed:2008}. Complex network analysis
tools have also been used to investigate the structure of energy
landscapes of various systems of interest, such as Lennard-Jones
atoms, proteins, or spin glasses, among others
\cite{Cieplak:1998_1,Cieplak:1998_2,Carmi:2009,Seyed:2008,Doye:2005,network_as_a_tool_1,network_as_a_tool_2,network_as_a_tool_3,network_as_a_tool_4,network_as_a_tool_5}.
The energy of a minimum and its degree (i.e., the number of other
minima which can be reached from this minimum) have been shown to be
correlated, as well as the barriers to overcome to escape from a
minimum. In particular, a logarithmic dependence of the energy of a
minimum on its degree has been exhibited, as well as energy barriers
increasing as a (small) power of the degree of a node
\cite{Carmi:2009,Doye:2002,Seyed:2008}. No systematic study of these
issues has however been performed, and most investigations have been
limited to relatively small systems because of computational limitations.

Most importantly, the investigations cited above have focused on the
topology of the network of minima, conceived as a tool to characterize
the energy landscape. The structure of a network has however a deep
impact on the properties of the dynamical processes which take place
on it \cite{Barrat:2008}. It seems thus adequate to put to use the
tools and techniques developed for the analysis of dynamical processes
on networks to achieve a better understanding of how the energy
landscape structure, represented as a network, affects the system
performing a random walk in it, and how the onset of glassy dynamics
can be described in this way in a general framework. In a previous
paper \cite{Baronchelli:2009}, we have made a first step to fill this
gap by focusing on the trap model put forward in Ref.
\cite{Bouchaud:1992}. In this paper, we generalize our approach to
more involved transition rates between energy minima. We show how the
heterogeneous mean-field (HMF) theory \cite{Barrat:2008,DorogoRev} can
be used in this context to highlight the connection between the
topological properties of the network of minima and the dynamical
exploration of these minima.  We show in particular that the
relationship between energy and degree of the minima is a crucial
ingredient for the existence of a transition and the subsequent glassy
phenomenology.  Our results shed light on the empirically found
relationship between the energy of a local minimum and its degree, and
we hope that they will stimulate more systematic investigations on
this issue.

We have organized our paper as follows: In Section~\ref{sec:model} we
define our model of energy landscape dynamics as a random walk on a
complex network. Different physical transition rates
are proposed, and the corresponding numerical
implementation is discussed. In Section~\ref{sec:formalism} we present a
theoretical analysis based on the heterogeneous
mean-field approximation for dynamical processes on complex networks. This
formalism is applied in Section \ref{sec:general-formalism}, where
general analytical approximate expression are presented for the main
quantities characterizing the glassy transition and dynamics. These
expressions are applied to the different physical transition rates
considered in Section~\ref{sec:appl-phys-trans}, where checks against
numerical simulations are also presented. In Section~\ref{sec:barriers} we
discuss the relation between energy basins and energy
barriers. Finally, in Section~\ref{sec:conclusions} we present our
conclusions.

\section{Random walk models on complex energy landscapes}
\label{sec:model}

\subsection{Definition}
\label{sec:definition}

\begin{figure}[t]
  \begin{center}
    \includegraphics*[width=9cm]{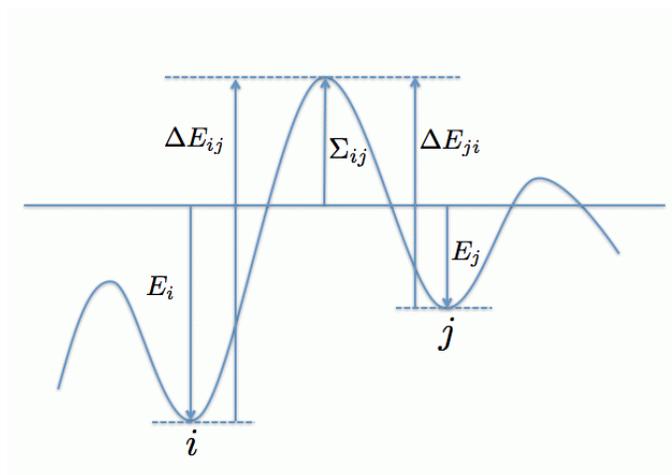}    
    \caption{Potential energy landscape description. Energies $E_i$
      are measured positive downwards. Energy gaps $\Sigma_{ij}$ are
      defined positive upwards.}
    \label{landscape}
  \end{center}
\end{figure}

We consider a network of $N$ nodes, in which each vertex $i$
corresponds to a minimum in the energy landscape, and a link is drawn
between two minima $i$ and $j$ if the system can jump directly from
$i$ to $j$. To each node $i$ is associated the energy $-E_i$ of the
corresponding minimum (energies are defined from a reference level, in
such a way that $E_i>0$ for all $i$). Moreover, an energy gap
$\Sigma_{ij}$ is associated to the edge between vertices $i$ and $j$,
as depicted in Fig.~\ref{landscape}: $\Sigma_{ij}$ is a symmetric
function, such that the energy barrier that must be overcome to jump
from vertex $i$ to vertex $j$ can be written as $\Delta E_{ij} = E_i +
\Sigma_{ij}$ and, analogously, $\Delta E_{ji} = E_j +
\Sigma_{ij}$. Obviously, we will have in general $\Delta E_{ij} \neq
\Delta E_{ji}$.

The system under investigation is pictured as a walker
exploring the network through a biased random walk. The rate
(probability per unit time) $r_{i \to j}$ to go from vertex $i$ to
vertex $j$ depends a priori on the energy at vertices $i$ and $j$
and/or the energy barrier between $i$ and $j$ that must be
overcome. The random walk model is defined in discrete time $t$ as
follows:
\begin{itemize}
\item At time $t$, the walker is in  vertex $i$.
\item It chooses at random a neighbor of $i$, namely $j$.
\item With a probability $r_{i \to j}$, that depends on the energy
  $E_i$ and/or on the energy barrier $\Delta E_{i j}$, the walker hops
  to vertex $j$.
\item Time is updated $t \to t+1$.
\end{itemize}

The relationship between the probabilities $r_{i \to j}$ and the
energy and energy barriers can be of different forms. In usual
unbiased random walks, $r_{i \to j}$ is a constant independent of both
$i$ and $j$ \cite{Noh:2003}. As a first step to introduce a dependence
on the nodes, a possible approach is that in which the energy barriers
depend only on the local minima themselves, i.e. we consider
$\Sigma_{ij}=0$.  For example, in the Bouchaud trap model considered
in Ref.~\cite{Bouchaud:1992}, the probability to exit from a trap is
just an Arrhenius law depending only on the departing trap's depth,
namely
\begin{equation}
  \label{eq:rate_traps}
  r_{i \to j}^{traps}= r_0 e^{-\beta E_i},
\end{equation}
where $\beta=1/T$ is the inverse temperature and $r_0$ is a constant
that determines a global timescale.  Other possible definitions
include the Metropolis one
\begin{equation}
  \label{eq:rate_Metropolis}
  r_{i \to j}^{Metropolis}= r_0 \min \left(1, e^{\beta(E_j-E_i)} \right),
\end{equation}
and the Glauber rate
\begin{equation}
  \label{eq:rate_Glauber}
  r_{i \to j}^{Glauber}= \frac{r_0}{1+ e^{-\beta(E_j-E_i)}}.
\end{equation}
We note that the rates considered in the Bouchaud trap model are quite
different from the Metropolis or Glauber rates. Indeed, while the
former depends only on the depth of the originating trap, the latter
depend also on the energy of the arriving vertex. This translates in
the fact that, in the limit of zero temperature, the dynamics of the
Bouchaud trap model is frozen for any $E_i$, while Metropolis and
Glauber dynamics still allow jumps to lower energy minima
\cite{Barrat:1995}. Within an even more realistic representation of
glassy dynamics, one can also contemplate the case $\Sigma_{ij}\neq0$,
allowing for the transition rates to depend explicitly on the energy
barriers between adjacent minima.  As a paradigm of this choice, we
propose a rate of the Arrhenius form
\begin{equation}
  \label{eq:rate_barriers}
  r_{i \to j}^{barriers}= r_0 e^{-\beta \Delta E_{ij}},
\end{equation}
which acts a straightforward generalization of the local transition rate
(\ref{eq:rate_traps}).

The case of rate (\ref{eq:rate_traps}) (local trapping) was studied in
a previous publication \cite{Baronchelli:2009}. In the following, we
will consider in turn non-local rates (\ref{eq:rate_Metropolis}),
(\ref{eq:rate_Glauber}) and (\ref{eq:rate_barriers}) and discuss the
fundamental differences due to the introduction of energy barriers in
the model.

We emphasize that our model differs both from usual unbiased random
walks, as the local energy determines the transition rates, and from
mean-field trap models in which jumps between any pair of energy
minima are a priori possible: Here, the system can jump only between
neighboring nodes. The dynamical evolution depends therefore both on
the network topology and on the energies associated to the nodes.

\subsection{Numerical implementation}
\label{sec:numer-impl}

To implement numerically the random walk, it is convenient to resort
to the techniques developed for general diffusion processes on complex
weighted networks \cite{rw_weight}. The main advantage of this method
is to avoid rejection steps, thus improving dramatically the
computational efficiency \cite{Bortz:1975,Krauth}. Therefore, at each
simulation step the random walker sitting at node $i$ selects a
neighbor $j$ with probability $r_{i \to j} / \sum_j r_{i \to j}$,
where the sum in the normalizing factor is extended to all of $i$'s
neighbors. As the walker hops on node $j$, the physical time is
incremented by an interval $\Delta t$ drawn from the exponential
distribution $P(\Delta t) = 1/\overline{\Delta t} \exp (- \Delta t /
\overline{\Delta t})$, where $\overline{\Delta t} =k_i / \sum_j r_{i
  \to j}$ is the inverse of the average escape rate out of node
$i$. In this way the simulation time is disentangled from the
physical time and the latter has no impact on the simulation
efficiency. No matter how much physical time a walker spends in a
node, from the simulation time point of view it is always just a time
step.

The network substrates on which we will focus are scale-free networks
with a degree distribution of the form $P(k) \sim k^{-\gamma}$ and $ 2
< \gamma \leq 3$. We will generate them with the Uncorrelated
Configuration Model (UCM) \cite{Catanzaro:2005}, that allows us to
tune the degree distribution to the desired form and prevents the
formation of degree-degree correlations. Networks are therefore
generated as follows:  A number of stubs (or semi-links) extracted
from the desired final degree distribution is assigned to each
node. Stubs are then randomly paired to form links between nodes, with
the prescription that multiple links as well as self-loops must be
avoided. A minimal degree $m$ is fixed a priori. To avoid spurious
effects due to the possible presence of tree-like structures
\cite{baronchelli2008random} it is convenient to adopt $m>2$. We will
choose $m=4$ in all of our simulations. So far the algorithm
coincides with the one of the Configuration Model \cite{molloy95}, but
the UCM introduces moreover a cutoff to the degree distribution, $k_c =
N^{1/2}$, which avoids the formation of degree correlations by limiting
the size of the hubs \cite{Catanzaro:2005}.

\section{Heterogeneous mean-field theory}
\label{sec:formalism}

In order to gain analytical understanding of the role of the different
transition rates in the corresponding glassy dynamics, we apply a
standard heterogeneous mean-field (HMF) formalism
 \cite{Barrat:2008,DorogoRev} . The basic tenet of HMF is the assumption
that all the dynamical properties of a vertex depend only on its
degree. Vertices are thus grouped into classes according to their
degree, and vertices with the same degree are treated as
equivalent. This approximation is consistent with previous findings
that have uncovered the correlations between the energy of a local
minimum and the degree of the corresponding node in the network
\cite{Doye:2002}. We therefore make the assumption that there exists a
relationship $E_i = h(k_i)$ where the function $h(k)$ is a
characteristic of the model. This also means that the distributions of
energies $\rho(E)$ of the system's landscape, and the degree
distribution $P(k)$ of the corresponding network are linked through $h$. In the
same spirit, we make the further assumption that the energy
gap between minima $i$ and $j$ depends only on the degrees of 
$i$ and $j$, i.e., that it
can be written as $\Sigma_{i,j} = \sigma(k_i, k_j)$,
where $\sigma(k, k')$ is a symmetric function of $k$ and $k'$.

Under the HMF approximation the dynamics will thus focus on the transitions
between different degree classes. The rate to go from
a vertex $k$ to a vertex $k'$ can be written as
\begin{equation}
  W_{k k'} = P(k'|k) r(k \to k').
  \label{eq:weighted_propagator}
\end{equation}
The function $P(k'|k)$, defined as the conditional probability that a
vertex of degree $k$ is connected with another vertex of degree $k'$
\cite{alexei}, takes into account the topological features of the
network, by gauging the probability of selecting a vertex $k'$ as
neighbor of $k$. The function $r(k \to k')$ measures the
rate of jumping from a vertex of degree $k$ to 
a vertex of degree $k'$ (given that they are connected
by an edge), and depends on $k$ and $k'$ through the rates $r_{i\to
  j}$ and the functions $h$ and $\sigma$. Obviously, the
rate $r(k \to k')$ is not in general a symmetric function of $k$ and
$k'$. It is worth noting that, apart from a normalization, 
Equation~(\ref{eq:weighted_propagator}) is simply the so-called
weighted propagator describing the probability that a node in class $k$
interacts with a node in class $k'$ \cite{rw_weight}. We also note that
the rates  $r(k \to k')$ depend on the inverse temperature $\beta$
through the microscopic rates $r_{i \to j}$.

\section{General HMF formalism}
\label{sec:general-formalism}

In this section, we apply the HMF theory to compute different
quantities relevant for the characterization of the dynamics of
a random walk in a complex energy landscape represented in terms of a
network of minima.

\subsection{Occupation probability}
\label{sec:occup-prob}

The description of a random walk dynamics starts from the
occupation probability $P(k, t_w)$, defined as the probability for the
walker to be in {\it any} node of degree $k$ at a time $t_w$. Its time
evolution can be easily represented in terms of a master equation of
the form
\begin{equation}
  \label{rate_scalar}
  \frac{\partial P(k,t_w)}{\partial t_w} \equiv \dot P(k, t_w) =
  -\sum_{k'} W_{k k'}P(k, t_w)+ 
    \sum_{k'} W_{k' k} P(k',t_w). 
\end{equation}
Upon describing the state at time $t_w$ with the row vector ${\bf
  P}(t_w)=\left\{P(1,t_w),P(2, t_w),\!...\!, P(k_c, t_w)\right\}$,
where $k_c$ is the cutoff or largest degree in the network, Equation
(\ref{rate_scalar}) can be rewritten in vector form as
\begin{equation}
  \label{rate_vectorial}
  \dot {\bf P}(t_w)=-{\bf P}(t_w) L \ ,
\end{equation}
where the matrix $L$, with elements
\begin{equation}
L_{k'k}=\left(\delta_{k'k}\sum_l W_{kl} - W_{k'k}\right) \ ,
\end{equation}
is a generalization of a Laplacian
matrix to the case of directed weighted graphs. The matrix elements satisfy
\begin{equation}\label{lin_dep}
L_{k'k'}=\sum_{k,k\neq k'} L_{k'k} \ ,
\end{equation}
which ensures conservation of probability and states that the columns of
$L$ are not linearly independent.
The real part of every eigenvalue of $L$ is non-negative \cite{AGA-05}.
As a consequence, all solutions of Eq. (\ref{rate_vectorial}), which
can be formally written as
\begin{equation}
  {\bf P}(t_w)={\bf P}(t_0)e^{-L(t_w-t_0)}  \ ,
\end{equation}
are stable according to Lyapunov criteria. In particular, since $\det
L=0$, $L$ always has the eigenvalue $0$, which corresponds to a
constant solution of the problem. At this point one can proceed
in close analogy with discrete-time regular Markov chains
\cite{meyer}. By making the assumption that the matrix $W_{k'k}$ is
non-negative and irreducible (indeed it is for every choice of
$r(k'\rightarrow k)$ in the following), we can prove that the $0$
eigenvalue of $L$ has algebraic multiplicity $1$. Hence, the
stationary solution of Eq. (\ref{rate_vectorial}) is unique.

\subsubsection{Steady state}

In order to calculate the steady solution ${\bf P}^{\infty}$ in the
limit $t_w \to \infty$, one can impose $\dot {\bf P}(t_w)=0$. This
leads to the condition
\begin{equation}
  \label{homogeneous_v}
  {\bf P}^{\infty}L=0,
\end{equation}
so that we are left with the task of finding the left nullspace of
$L$. Eq. (\ref{homogeneous_v}) is a homogeneous system of algebraic
linear equations. It admits non-trivial solutions since $\det(L)=0$.
In our case,  the solution to
(\ref{homogeneous_v}) can be easily found by imposing the detailed balance
condition. Namely, writing Eq.~(\ref{homogeneous_v}) as
\begin{equation}
  \sum_{k'} \left[-W_{k k'}P^\infty(k)+ W_{k' k}
    P^\infty(k')\right]=0,  
  \label{eq:1}
\end{equation}
we can obtain a solution by imposing that the terms inside the
summation in Eq.~(\ref{eq:1}) cancel individually, that is
\begin{equation}
  W_{k k'}P^\infty(k)= W_{k' k} P^\infty(k'), \qquad \forall k, k'.
\end{equation}
Substituting the form of $W_{k k'}$, we obtain
\begin{equation}
  \frac{P^\infty(k)}{P^\infty(k')} = \frac{W_{k' k}}{W_{k k'}} =
  \frac{P(k|k') r(k' \to k)}{P(k'|k) r(k \to k')} 
  = \frac{ k P(k)}{k' P(k')} \frac{r(k' \to k)}{r(k \to k')},
  \label{eq:2}
\end{equation}
where in the last step we have used the degree detailed balance
condition $k P(k) P(k'|k) = k' P(k') P(k|k')$ which simply expresses
that the number of edges from a node of degree $k$ to a node of degree
$k'$ is equal to the number of edges from a node of degree $k'$ to a
node of degree $k$ \cite{marian1}. From Eq.~(\ref{eq:2}), we see that
its right-hand-side must be expressible as a simple ratio of a
function of $k$ over a function of $k'$. A general way to obtain this
is to impose a coarse-grained rate $r(k'\rightarrow k)$ taking the
general form
\begin{equation}
  r(k'\rightarrow k)=f(k')g(k)s(k',k).
  \label{eq:3}
\end{equation}
In other words, we assume that the rate $r(k'\rightarrow k)$ can be
written as the product of a function of $k'$, a function of $k$, and a
symmetric function $s(k',k)=s(k,k')$ (where $k$ and $k'$ need not be
separable). We will see later that all the rates $r_{i\to j}$ defined in 
Sec. \ref{sec:definition} (traps, Glauber, Metropolis, and
energy barriers) can be written in such a form.
The stationary solution is then given by
\begin{equation}
  \label{solution}
  P^\infty(k) =\frac{1}{\mathcal{Z}}kP(k)g(k)/f(k)
\end{equation}
where ${\mathcal{Z}}$ is a normalizing constant determined by the
condition $\sum_k P^\infty(k)=1$. Such a
solution is unique, as proven above. Interestingly, the
symmetric function $s(k',k)$ does not enter the steady solution,
although it will play a role in affecting the transient
behavior, as we will see in the next sections.

\subsubsection{Glassy phase}

The steady state solution found above for the
occupation probability is defined if and only if the
normalization constant
\begin{equation}
  \mathcal{Z} = \sum_k kP(k)g(k)/f(k).
  \label{eq:4}
\end{equation}
is finite. When this condition is met, the random walker reaches an
equilibrium state with a distribution ${\bf P}^{eq} = {\bf
  P}^\infty$. On the other hand, whenever such condition is not met,
the random walker is unable to reach a steady state, i.e. the steady
solution to the rate equation does not correspond to any physical
steady state in equilibrium ${\bf P}^{eq}$. We identify this region of
the phase space with the glass phase for our random walker \cite{Bouchaud:1992}.

The functions $f$ and $g$ depend on the temperature, on the precise
dynamics chosen (traps, Glauber, Metropolis), and encode the
relationship $h$ between energy and degree of the minima. The degree
distribution moreover enters explicitly the expression $\mathcal{Z}$.
As the various parameters of the model are changed, it is thus a
priori possible to go from one phase in which $\mathcal{Z}$ is finite
to one in which $\mathcal{Z}$ diverges. In a physical system in
particular, the control parameter is usually the temperature, while
the topology of the network of minima and the function $h$ are
given. It is then clear from Eq. (\ref{eq:4}) that the presence or
absence of a finite glass transition temperature $\beta_c$, such that
$\mathcal{Z}$ becomes infinite for $\beta \geq \beta_c$, depends on
the interplay between the topology of the landscape network (as
determined by $P(k)$) and the functions $f$ and $g$.  Interestingly,
at this mean-field level, the existence of a transition does not
depend on the network degree correlations, since the conditional
probability $P(k'|k)$ do not enter Eq. (\ref{eq:4}).

Let us consider for instance a network of minima with a heavy-tailed
degree distribution such as $P(k) \sim k^{-\gamma}$. A transition
between a finite and an infinite $\mathcal{Z}$ can be observed if and
only if $g(k)/f(k)$ behaves at large $k$ in the form $\sim k^{a}$
where the exponent $a$ depends on the temperature, and can take values
smaller or larger than $\gamma-2$ depending on the temperature.
Another example is given by a stretched exponential form for $P(k)$,
$P(k) \sim e^{-b k^{a}}$, in which case a transition is observed if
and only if $g(k)/f(k)$ is of the form $e^{b' k^{a}}$, with $b'$ a
function of the temperature (the transition is then given by
$b'(\beta_c)=b$).

\subsubsection{Glassy dynamics}
\label{sec:glassy-dynamics}

In any finite system, unless the product function $g(k)/f(k)$
exhibits some sort of singularity, the normalization constant
Eq.~(\ref{eq:4}) is finite and the steady state distribution
$P^\infty(k)$ exists, the occupation probability $P(k, t_w)$
converging to it after an equilibration time, i.e.
\begin{equation}
  \label{eq:5}
  \lim_{t_w \to \infty} P(k, t_w) = P^\infty(k).
\end{equation}
The corresponding thermalization of the occupation probability occurs
in a way depending on the function $h$. Shallow energy minima are
indeed explored first, while deep traps (large $E$) are visited at
larger times \cite{Bouchaud:1992,Barrat:1995}. If $h$ is a growing 
function of $k$, as indeed found empirically \cite{Doye:2002}, small
degree nodes correspond to shallow minima, and deeper minima are
associated to larger nodes. The evolution of $P(k, t_w)$ takes then
place in a hierarchical fashion: The small degree region equilibrates
first, and progressive equilibration of larger degree regions takes
place at larger times. In this respect, we obtain a strong difference
between the biased random walk that the glassy system experiences and
usual diffusion processes corresponding to unbiased random walks,
which visit first large degree vertices and then cascade down towards
small degree nodes \cite{Noh:2003,Barthelemy:2005,Barrat:2008}, in the
present case we observe an ``inverse cascade process'' from small
vertices to hubs.

We have found in \cite{Baronchelli:2009} that, in the case of a random
walk among traps, this hierarchical thermalization is summarized in a
scaling form for $P(k, t_w)$, which can be written as
\begin{equation}
  \label{eq:6}
  P(k, t_w) = k_w(t_w)^{-1} \mathcal{F}\left(\frac{k}{k_w(t_w)}\right),
\end{equation}
where $k_w(t_w)$ represents the maximum degree of the vertices
equilibrated up to time $t_w$, and $\mathcal{F}(x)$ interpolates
between $P^\infty(x)$ at small $x$ and the short time form of $P(k,
t_w)$ which is proportional to $k P(k)$. We will see in the next
section that a similar scaling is obeyed for other transition rates.
In general, for the glassy dynamics, the functional form of $k_w(t_w)$
can moreover be obtained through the following argument: The total
time $t_w$ can be written as the sum of the trapping times spent in
the vertices that have been visited since the beginning of the
dynamics. Trapping times increase with the depth of the minimum, hence
with the degree (we are still considering the case of an increasing
function $h(k)$), and, in the glassy phase, the consequence is that
the sum of trapping times is dominated by the vertex with the largest
degree visited up to that point, namely $k_w$. Moreover, the average trapping
time $\tau_k$ at a given vertex $k$ can be estimated as the inverse of
the average escape rate $r_k$ from that vertex:
\begin{equation}
  \label{eq:7}
  \frac{1}{\tau_k} = r_k = \sum_{k'} W_{kk'}=\sum_{k'}P(k'|k)r(k\to
  k') = \sum_{k'} P(k'|k) f(k) g(k') s(k,k'). 
\end{equation}
We can therefore estimate $k_w(t_w)$, the typical degree
up to which nodes are ``equilibrated'' at time $t_w$, 
by approximating $\tau_{k_w} \sim t_w$, and solving the equation
\begin{equation}
  \label{eq:8}
  t_w  
= \frac{1}{f(k_w)}
  \frac{1}{\sum_{k'}P(k'|k_w)g(k') s(k', k_w)} 
\end{equation}
to obtain $k_w$ as a function of $t_w$. Note
that the result depends here on the function $s(k,k')$ and not only
on $f$, $g$, $h$ and $P(k)$.

\subsection{Average escape time}
\label{sec:escape}

The properties of the system can be further quantified by
measuring the average time $t_{esc}(t_w)$ required by the random
walker to escape from the vertex it occupies at time $t_w$
\cite{Baronchelli:2009}. For small waiting times $t_w$, $t_{esc}$
increases as a result of the transient equilibration of $P(k,t)$.  For
large $t_w$, such that $P(k,t_w)$ is close enough
to the equilibrium $P^\infty$, $t_{esc}$ can be
calculated instead as the average
 \begin{equation}\label{eq:t_esc}
 t_{esc}(t_w\to\infty)=\sum_k P^\infty(k) \tau_k = 
\frac{1}{\mathcal{Z}} \sum_k kP(k)  [g(k)/f(k)] \tau_k
 \end{equation}
where $\tau_k = 1/r_k$ is the inverse of the equilibrium escape rate, cf
Eq. (\ref{eq:7}), yielding
\begin{equation}
  \label{eq:10}
  t_{esc}(t_w\to\infty) = \frac{1}{\mathcal{Z}} \sum_k 
    \frac{k P(k) g(k)}{f(k)^2 \sum_{k'} P(k'|k) g(k') s(k',k)}.
\end{equation}
Most interestingly, the explicit form of the average escape time
$t_{esc}$ depends explicitely on the
symmetric function $s(k,k')$ as well as on the network degree
correlations, as expressed by the conditional probability
$P(k'|k)$.

\subsection{Average rest time}
\label{sec:average-rest-time}

Let us go back to the issue of the existence of a glass transition in
the model. We first recall the phenomenology of the fully connected
trap model, with transition rates $r_{i \to j}= r_0 e^{-\beta E_i}/N$
for any $i$ and $j$, where the energies $E_i$ are random numbers
extracted from a distribution $\rho(E)$
\cite{Bouchaud:1992,Monthus:1996}. As all traps are connected with
each other, all traps are equiprobable after a jump, so that the
probability for the system to be in a trap of depth $E$ is simply
$\rho(E)$, and the average rest time spent in a trap is $\langle \tau
\rangle= \int \rho(E) e^{\beta E} dE$. A transition between a high
temperature phase and a glassy one is thus obtained if and only if,
when $\beta$ increases, $\langle \tau\rangle$ is finite at small $\beta$ and
diverges at a finite $\beta_c$. Such a phenomenology is obtained if
and only if $\rho(E)$ is of the form $\exp(-\beta_c E)$ at large $E$
(else the transition temperature is either $0$ or $\infty$), and the
transition temperature is then $T_c=1/\beta_c$ \cite{Bouchaud:1992}.

In the present case of a nework of minima, the average rest time that
the walker spends in a minimum is
\begin{equation}
  \label{eq:9}
  \langle \tau\rangle=\left\langle \frac{1}{r_k}\right \rangle_h,
\end{equation}
where the symbol $\langle ... \rangle_h$ refers to the average
performed over the measure $P_h(k)$, which represents the probability
that the walker is in any vertex of degree $k$ after a hop. Note that
we disregard here the {\em physical} time, which is the sum of times
spent in the various minima, but consider only the number of hops
between minima. In the case of the traps model, $P_h$ is simply given
by the probability to be in a node of degree $k$ after a hop in a
random walk, i.e. by $kP(k)/\langle k \rangle$
\cite{Baronchelli:2009}, since the transition rates do not depend on
the arrival node. In a non-local trapping model instead, we need to
write a master equation of the form
\begin{equation}
  \dot{P}_h(k)=-P_h(k)+\sum_{k'}\mathcal{W}_{k'k}P_h(k'),
\end{equation}
where the matrix
$\mathcal{W}_{k'k}=W_{k'k}/\sum_lW_{k'l}=W_{k'k}/r_{k'}$ is now
stochastic and the derivative is intended with respect to the number of
hops. In the long time limit, we impose 
$\dot{P}_h(k)=0$ and calculate $P_h(k)$ as
we did for $P^\infty(k)$, imposing the detailed balance
condition, and obtaining
\begin{equation}
P_h(k) = \frac{1}{\mathcal{I}} k P(k) [g(k)/f(k)]  r_k \ ,
\end{equation}
where $\mathcal{I}$ is a normalization factor, given by
\begin{equation}
  \mathcal{I}=\sum_{k}\sum_{l}kP(k)P(l |k) g(k)g(l)
  s(k,l).
\end{equation}

We finally obtain for the average $\langle \tau \rangle$:
\begin{equation} \label{tau_zi}
  \langle \tau \rangle= \sum_k P_h(k) /r_k =\frac{\mathcal{Z}}{\mathcal{I}},
\end{equation}
where $\mathcal{Z}$ is the normalization factor of $P^\infty(k )$
defined in Eq.~(\ref{eq:4}).
As for the average escape time,
the average rest time $\langle \tau \rangle$ thus depends on
all the parameters of the system, including  
the network's degree correlations and the symmetric function $s$.

\section{Application to physical transition rates}
\label{sec:appl-phys-trans}

In this section, we apply the general HMF results obtained
in the previous section to physical transition probabilities between
local minima given by the trap model, Glauber, Metropolis and
barrier-mediated rates. We will focus for definiteness on scale-free
networks characterized by a power-law degree distribution $P(k)\sim
k^{-\gamma}$ with $2<\gamma\leq3$, which turns out to be the interval
reported in the literature \cite{Carmi:2009,Doye:2002}.

Let us first consider the explicit form of the transition
rates in each case, to show that 
they can be cast in the form given by Eq.~(\ref{eq:3}). In the case of the trap
model, the rate to jump from a vertex $k$ to a vertex $k'$ is simply
$r(k \to k') = r_0 e^{-\beta E_k} = r_0 e^{-\beta h(k)}$, where we recall
that $h(k)$ gives the depth of a node of degree $k$: it depends
only on the degree of the starting node, and not on the node reached
after the jump. We can therefore use
\begin{equation}
  \label{eq:11}
  \mbox{(Trap model)} \quad f(k) = e^{-\beta h(k)}, \;\; g(k) = 1,
  \;\; s(k,k') =r_0.
\end{equation}
The Glauber rate can be written as
\begin{equation}
  r(k \rightarrow k' )=r_0 \frac{e^{\beta h(k')}}{  e^ {\beta h(k)
    }+e^{\beta h(k')} },
\end{equation}
leading to
\begin{equation}
  \label{eq:12}
  \mbox{(Glauber)} \quad f(k) = 1, \;\; g(k) =  e^{\beta h(k)},
  \;\; s(k,k') =r_0 \frac{1}{e^{\beta h(k) }+e^{\beta h(k')}}.
\end{equation}
The Metropolis transition, on its turn, reads
\begin{equation}
  r(k \rightarrow k')=r_0 \min\left[1,e^{\beta (h(k')- h(k))}\right].
\end{equation}
Since, for positive $a$, $\min(1,b/a)=\min(a,b)/a$,
we can choose
\begin{equation}
  \label{eq:13}
  \mbox{(Metropolis)} \;f(k) = e^{-\beta h(k)}, \; g(k) = 1 ,
  \; s(k,k') =r_0 \min(e^{\beta h(k)}, e^{\beta h(k')}).
\end{equation}
Finally, in the presence of energy barriers, the transition rate reads
\begin{equation}
  r(k \to k')= r_0 e^{-\beta(h(k)+ \sigma(k,k'))},
\end{equation}
where $\sigma(k,k')$ is a symmetric function of its arguments, so that
we can use
\begin{equation}
  \label{eq:14}
  \mbox{(Barriers)} \;f(k) = e^{-\beta h(k)}, \; g(k) = 1 ,
  \; s(k,k') =r_0  e^{-\beta \sigma(k,k')}.
\end{equation}

\subsection{Steady state and the glass transition temperature}

Interestingly, for all the  transition rates considered above, the
product of the functions $1/f$ and $g$, which controls 
the existence of the steady state
solution of the occupation probability, takes the form
\begin{equation}
  \label{eq:15}
   g(k)/f(k) \equiv e^{\beta h(k)}.
\end{equation}
The normalization constant $\mathcal{Z}$ defined in Eq.~(\ref{eq:4})
can thus be written as
\begin{equation}
  \label{eq:16}
  \mathcal{Z} = \sum_k k P(k) e^{\beta h(k)}.
\end{equation}
For a power-law degree distribution $P(k) \sim k^{-\gamma}$, 
a finite glass transition
temperature is then obtained if and only if $h$ is of the form
\begin{equation}
  \label{eq:17}
  h(k)=E_0 \log (k) \ ,
\end{equation}
which is precisely what has been found, in conjunction with a 
scale-free degree distribution, in Ref. \cite{Doye:2002}.
$\mathcal{Z}$ is then indeed given by a sum of terms of the form
$k^{1-\gamma+\beta E_0}$, which converges if and only if 
\begin{equation}
  \beta E_0 -\gamma < -2 \ .
  \label{eq:18}
\end{equation}
In other words, a transition between a high temperature phase 
in which $P^\infty(k)$ exists and a low temperature glassy phase
is obtained at the critical temperature~\cite{Baronchelli:2009}
\begin{equation}
  \label{eq:tc}
  T_c =  \frac{1}{\beta_c}= \frac{E_0}{\gamma - 2} \ .
\end{equation}

Quite noticeably, the existence of a transition at a finite
temperature, as well as the value of this temperature, does not depend
on the form of the transition rates between the local minima, but only
on the existence of a particular interplay between the topology of the
network of minima and the relationship between energy and degree in
this network, as determined by the function $h$. 
We emphasize that this result is also independent
of the network degree correlations $P(k'|k)$, as already noted
in the previous section.

\subsection{The steady state and finite size effects}

\begin{figure}[t]
  \begin{center}
    \includegraphics*[width=9cm]{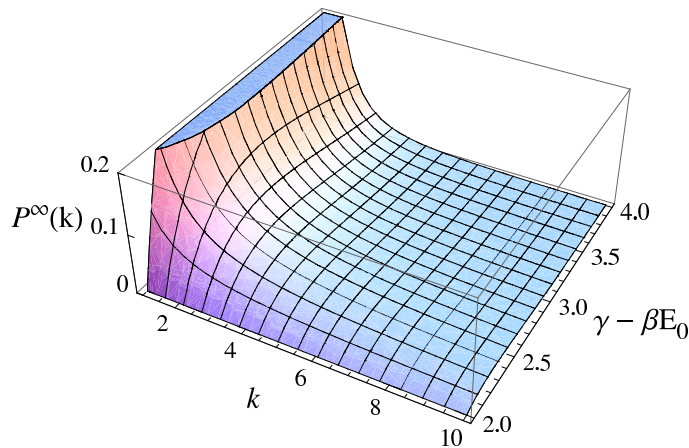}
    \caption{Equilibrium probability distribution $P^\infty(k)$ for
      the random walker to be in any node of degree $k$. For $\gamma
      -\beta E_0=2$ the system undergoes a transition to a glassy
      state.}
    \label{q_k}
  \end{center}
\end{figure}

\begin{figure}[t]
\begin{center}
 \includegraphics*[width=9cm]{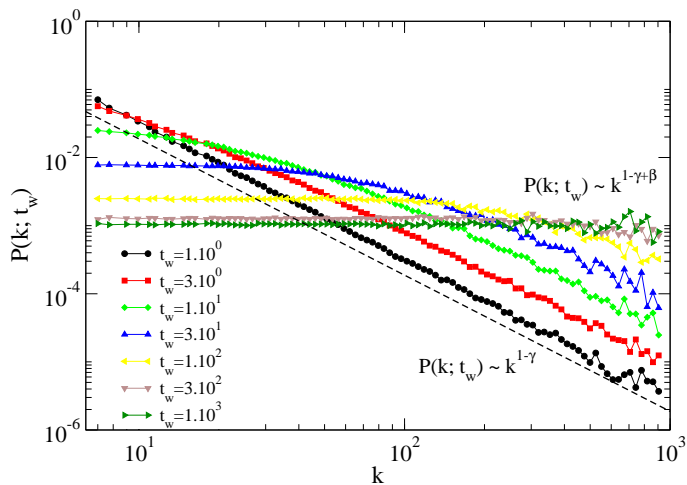}
\caption{Evolution towards equilibrium of the probability distribution
  $P(k; t_w)$ for Glauber dynamics. The distribution measured after a
  small waiting time $t_w$ is determined by a usual unbiased random
  walk behavior, i.e $P(k; t_w) \sim kP(k)$, while at larger times it
  relaxes to the equilibrium $P^\infty(k)$. The relaxation towards
  equilibrium starts from small degree nodes. Data refer to UCM
  networks with $N=10^6$ and $\gamma=3.0$ for $\beta=2$ ($E_0=1$):
  with these parameter values, for small times $P(k; t_w) \sim
  k^{-2}$, while at large enough times $P(k; t_w) \sim P^\infty(k)=const$.}
\label{q_k_sim}
\end{center}
\end{figure}

Let us focus on the case of a scale-free network of minima, with $P(k) \propto
k^{-\gamma}$ and $E_i=E_0 \log(k_i)$. For any of the rates
discussed above, the steady state measure, when it exists, is given by
\begin{equation}\label{eq:steady_state}
  P^\infty(k)=\frac{k^{1-\gamma+\beta E_0}}{\zeta(-1+\gamma -\beta E_0)}\;
  \mbox{for}\;\gamma-\beta E_0>2,
\end{equation}
where $\zeta$ is the Riemann $\zeta$ function. A plot of $
P^\infty(k)$ as a function of $k$ and $\gamma-\beta E_0$ is given in
Fig. \ref{q_k}, while data from simulations are reported in Fig.
\ref{q_k_sim} for the evolution of
$P(k,t_w)$ under Glauber dynamics. 
Above the transition ($\gamma-\beta E_0 > 2$), low-$k$ states (i.e., shallow
minima) are more probable. As the temperature decreases, $P^\infty(k)$ becomes less and less
peaked at low values of $k$, and large-$k$ states, which correspond
to lower energies, become more and more
probable.

In any finite system, the sum defining $\mathcal{Z}$, is finite
at any temperature as the degree distribution has a cut-off at a finite
$k_c$:
\begin{equation}
  \mathcal{Z} = \sum_k^{k_c} kP(k)g(k)/f(k).
\end{equation}
For instance, for $P(k) \propto k^{-\gamma}$, and with $h(k)=E_0
\log(k)$, the sum
\begin{equation}
  \sum_{k=1}^{k_c} k^{1-\gamma +\beta E_0}=H_{k_c}^{ (-1+\gamma-\beta
    E_0)}
  \label{eq:19}
\end{equation} 
is analytic in $\gamma-\beta E_0=2$ for any finite $k_c$.  Here
$H_{k_c}^{ (\alpha)}$ is the Harmonic Number of order
$\alpha$, which tends to $\zeta (\alpha)$
for $k_c \to \infty$.

The probability $P^\infty(k)$ is thus well defined
for every $\gamma-\beta E_0$ and for any finite system. 
In particular, performing a continuous degree
approximation in Eq.~(\ref{eq:19}), we can obtain an estimate of the
network size dependence of $\mathcal{Z}$ as
\begin{equation}
  \label{eq:20}
  \mathcal{Z} \sim \int^{k_c} k^{1-\gamma +\beta E_0} \;dk \sim
  \mbox{const} + k_c^{2-\gamma+\beta E_0}.
\end{equation}
For $\beta < (\gamma-2)/E_0$, $\mathcal{Z}$ tends to a constant
as the network size (and thus $k_c$) increases. On the other
hand, for $\beta > (\gamma-2)/E_0$, $\mathcal{Z}$ diverges as
$k_c^{2-\gamma+\beta E_0}$, i.e., as $N^{(2-\gamma+\beta E_0)/2}$
in uncorrelated scale-free networks, which obey $k_c \sim N^{1/2}$.

\subsection{Glassy dynamics}
\label{sec:glassy-dynamics-1}

At low temperatures, even for a finite system, the evolution of
$P(k; t_w)$ towards $P^\infty(k)$ is slow, as displayed in Fig. \ref{q_k_sim},
and an ageing regime takes place, in which the function $P(k,t_w)$
obeys the scaling form
\begin{equation}
   P(k, t_w) = k_w(t_w)^{-1}
  \mathcal{F}\left(\frac{k}{k_w(t_w)}\right),
  \label{eq:25}
\end{equation}
where the characteristic degree $k_w$ can be estimated from
Eq.~(\ref{eq:8}). In order to simplify its computation, we will consider
an uncorrelated network of minima, such that
$P(k'|k)=k'P(k')/\langle k\rangle$, and we will work in 
the continuous degree approximation,
using the normalized form $P(k) =
(\gamma-1)m^{\gamma-1} k^{-\gamma}$, where $m$ is the minimum degree
present in the network.

In the case of the Glauber dynamics, the escape rate can be expressed,
within the above approximations, as 
\begin{eqnarray}
  r_k &=& \frac{1}{\avk} \int_m^\infty k' P(k') r_0 \frac{e^{\beta
      h(k')}}{  e^ {\beta h(k) 
    }+e^{\beta h(k')} } \; dk' \equiv
  \frac{m^{\gamma-1}(\gamma-1)}{\avk} \int_m^\infty 
  \frac{z^{1+E_0 \beta-\gamma}}{z^{E_0 \beta} + k^{E_0 \beta}} dz
  \nonumber \\
  &=& \Gamma\left[1, \frac{\gamma-2}{\beta}, 1+\frac{\gamma-2}{\beta},
    -\left(\frac{k}{m}\right)^\beta \right],
 \label{eq:21}
\end{eqnarray}
where we have used the relation $h(k) = E_0 \ln k$ and where
$\Gamma[a,b,c,z]$ is the Gauss Hypergeometric function.  Using the
asymptotic expansion for $z\to 0$ \cite{abramovitz}, we obtain that
the leading behavior for large $k$ yields
\begin{eqnarray}
  r_k\sim \left\{
    \begin{array}{lr}
      k^{-\beta_cE_0} & \beta>\beta_c \\
      k^{-\beta E_0} & \beta<\beta_c \\
    \end{array}
  \right.,
  \label{eq:22}
\end{eqnarray}
which leads to
\begin{eqnarray}
  \tau_k = \frac{1}{r_k} \sim \left\{
    \begin{array}{lr}
      k^{\beta_cE_0} & \beta>\beta_c \\
      k^{\beta E_0} & \beta<\beta_c. \\
    \end{array}
  \right..
  \label{eq:23}
\end{eqnarray}
From here, using the relation $\tau_{k_w} \sim t_w$, we obtain
\begin{eqnarray}
  k_w \sim \left\{
    \begin{array}{lr}
      t_w^{1/(\beta_c E_0)} & \beta>\beta_c \\
      t_w^{1/(\beta E_0)} & \beta<\beta_c. \\
    \end{array}
  \right..
  \label{eq:24}
\end{eqnarray}

\begin{figure}[t]
  \begin{center}
    \includegraphics*[width=9cm]{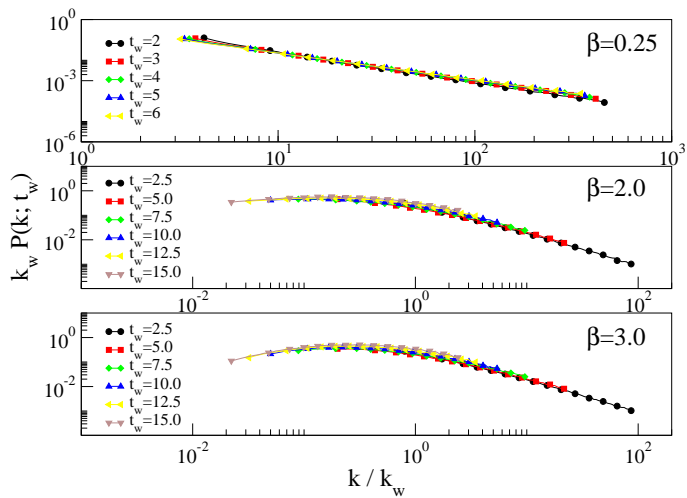}
    \caption{Data collapse for the time evolution of the occupation
      probability $P(k; t_w)$ at different temperatures (Glauber
      dynamics). Data refer to UCM networks with $N=10^6$ and
      $\gamma=2.5$, so that $\beta_c = \gamma-2=0.5$ (where we have
      taken $E_0=1$).  The top panel presents data for $\beta <
      \beta_c$ while both middle and bottom panels concern the low
      temperature case $\beta > \beta_c$. Accordingly, for the top
      panel we use $k_w \sim t_w^{1/\beta} \sim t_w^{4}$ for the
      rescaling, while for the central and the bottom ones it holds
      $k_w \sim t_w^{1/\beta_c} = t_w^{2}$ (see
      Eq.~(\ref{eq:24})). The curves corresponding to different $t_w$
      collapse well under this rescaling. Note that we use rather
      small values of $t_w$, because the equilibration time, defined
      by $k_w \sim k_c \sim N^{1/2}$, is $t_{eq}\sim N^{\frac{\beta_c}{2}} \simeq
      32$ for $\beta > \beta_c$ and $t_{eq}\sim N^{\frac{\beta}{2}}
      \simeq 6$ for $\beta < \beta_c$.  Each curve is obtained by
      averaging over $3 \times 10^6$ simulation runs.}
    \label{collapseP}
  \end{center}
\end{figure}

In Fig.~\ref{collapseP} we check the validity of Eqs.~(\ref{eq:25})
and (\ref{eq:24}) by performing a data collapse analysis for different
values of $t_w$. The curves obtained for different $t_w$ collapse indeed as predicted.

In the case of the Metropolis transition rates, a similar
analysis yields 
\begin{equation}
  r_k^{Metropolis}\propto \frac{1}{(\beta_c-\beta)E_0}\left(k^{-\beta
      E_0}-\frac{\beta}{\beta_c}k^{-\beta_c E_0}\right), 
\end{equation}
leading to the same asymptotic behavior as in Eq.~(\ref{eq:22}), and
therefore to the same scaling picture as for the Glauber rate.

As pointed out for the case of the traps model
\cite{Baronchelli:2009}, however, in finite systems the scaling
relations in Eq. (\ref{eq:24}) hold only as far as $k_w(t_w) < k_c$,
i.e. it exists an equilibration time $t_{eq}$, obtained by inverting
Eq. (\ref{eq:24}), above which the system has completely relaxed and
Eq.~(\ref{eq:25}) is no more valid. Finally, it is worth
stressing that, while for large temperatures the scaling exponent
relating $k_w$ and $t_w$ depends on the temperature, in the low
temperature phase it becomes independent, being just proportional to
the transition temperature. We note that this saturation of the
  exponent at $1/(\beta_c E_0)$ is very different from the
  phenomenology obtained in the trap model \cite{Baronchelli:2009},
  for which $k_w \sim t_w^{1/(\beta E_0)}$. An immediate consequence
  is that the equilibration time strongly depends on $\beta$, as
  $t_{eq} \sim k_c^{(\beta E_0)}$, for a system described by traps, but
  is given by $t_{eq} \sim k_c^{(\beta_c E_0)} \ll k_c^{(\beta E_0)}$ for
  any $\beta > \beta_c$ for Glauber and Metropolis rates.

\begin{figure}[t]
  \begin{center}
    \includegraphics*[width=9cm]{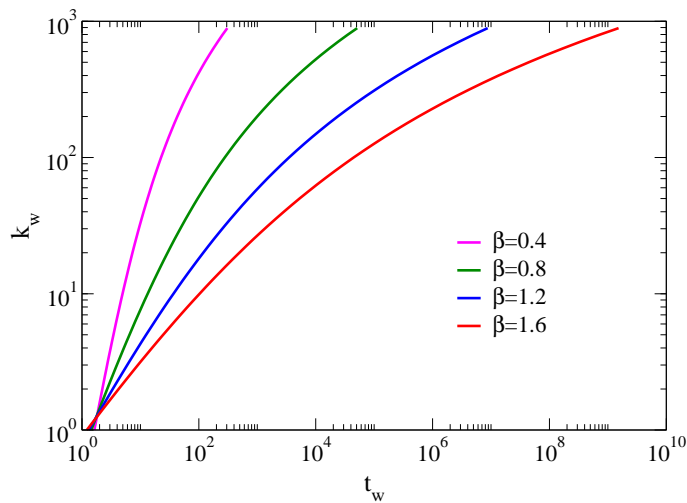}
    \caption{Maximum degree of equilibrated nodes up to time
      $t_w$ for barrier-mediated dynamics. Curves are obtained from the numerical inversion of
      Eq. (\ref{eq:8}). Values of parameters are: $\gamma=2.5$, $\beta_c=0.5$, $E_0= \epsilon=1$,
      $k_c=10^3$, $\mu=0.5$.  }
    \label{f:kw_barriers}
  \end{center}
\end{figure}

Contrarily to results for the glass transition temperature $T_c$ and
the steady state, the glassy dynamics for barrier-mediated rates does
not yield the same results as for Glauber and Metropolis rates, since
$r_k$ does depend on the symmetric function $\sigma (k,k')$. In
particular, we need here to choose a functional form for $\sigma$. 
We propose to use
\begin{equation}
  \sigma(k,k')=\sigma_0 (k^{\mu} + k'^{\mu}),
\end{equation}
which will be justified in Section \ref{sec:barriers}. In this case we obtain
\begin{equation}
  r_k^{barriers}\propto r_0k^{-\beta E_0} e^{-\beta \sigma_0 k^\mu}.
  \label{eq:29}
\end{equation}
In order to keep an interesting phenomenology,
the constant $\sigma_0$ cannot be chosen arbitrarily. 
If $\sigma_0$ were independent of the system size,
the escape rate would be dominated by the exponential at all
temperatures. This behavior would reflect the fact that in this case the
rate is suppressed in transitions involving nodes of
large $k$, eventually generating unphysically large barriers at large
$k_c$.  To prevent the system from building up infinite barriers, one
can impose
\begin{equation}
  E_0\ln k_c\propto  \sigma_0 k_c^\mu,
\end{equation}
such that the maximum barrier is always comparable to the lowest
energy minimum and neither term dominates the other. As a consequence,
we take
\begin{equation}
  \label{eq:size_depen}
  \sigma_0=\epsilon E_0 \frac{\ln k_c}{k_c^\mu}, 
\end{equation}
where $\epsilon$ is now constant and size independent. Contrarily to
the previous cases, $k_w(t_w)$ is hard to determine, as no explicit
inversion of Eqs. (\ref{eq:8}) and (\ref{eq:29}) can be provided for the range of
parameters of interest in our study. A numerical evaluation of
$k_w(t_w)$, is reported in Figure \ref{f:kw_barriers}. The maximum
degree of equilibrated vertices $k_w$ has an initial power law
increase in time, which is reminiscent of the local trapping
model. However, as larger degree nodes are equilibrated, exponential
barriers come into play and the hierarchical thermalization becomes
logarithmic in time.

\subsection{Average escape time}
\label{sec:average-escape-time}

\begin{figure}[t]
  \begin{center}
    \includegraphics*[width=9cm]{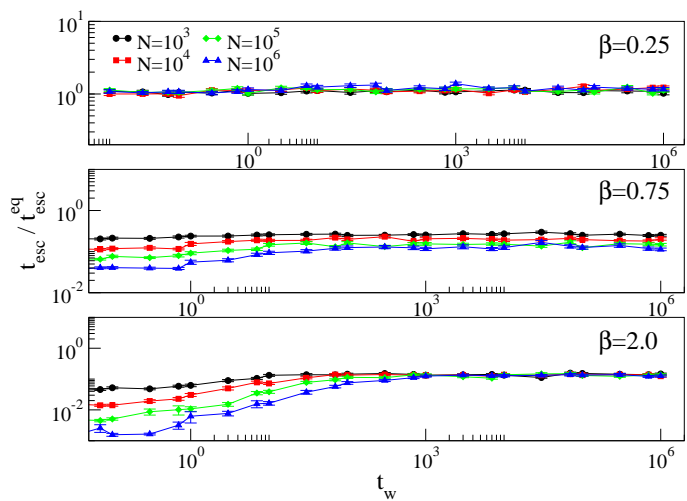}
    \caption{Rescaled average escape times (Metropolis dynamics). Data
      for UCM network with $\gamma=3.0$, so that $\beta_c=1.0$
      ($E_0=1$). The scaling forms of Eq.~(\ref{eq:escape}) produce a
      collapse of the curves concerning different systems sizes in the
      three regimes of high, intermediate and low temperature (top,
      center and bottom panels, respectively). The slightly worse
      collapse obtained for $\beta=0.75$ may be due to logarithmic
      corrections as $\beta$ is close to both $\beta_c$ and
      $2\beta_c$. Each point is averaged over $400$ simulation runs
      ($20$ runs on each of $20$ network realizations).  }
    \label{f:escape}
  \end{center}
\end{figure}

The average escape time, defined as the average time required by the
system to escape from the vertex it occupies, can be computed in the
long time limit from Eq.~(\ref{eq:t_esc}), as a function of the
average trapping time $\tau_k$ in vertices of degree $k$. From the
asymptotic expansions of $\tau_k$ in Eq.~(\ref{eq:23}), valid for
Glauber and Metropolis dynamics, evaluation of Eq.~(\ref{eq:t_esc})
allows us to observe that, whenever a finite $T_c$ exists,
$t_{esc}(t_w\to \infty)$ diverges at $2T_c$ in an infinite system, as
was already observed in the case of local trapping
\cite{Baronchelli:2009}. It is noticeable that the same divergence
temperature is obtained, as Eq. (\ref{eq:t_esc}) a priori involves the
network's degree correlations and the function $s$.  Within the
continuous degree approximation, the divergence of the escape time
with the system size follows the scaling laws
\begin{eqnarray}
  t_{esc}^{eq} \equiv t_{esc}(t_w\to \infty) \sim \int^{k_c}
  P^\infty(k) \tau_k \sim \left\{
    \begin{array}{lc}
      k_c^{\beta_c E_0} & \beta>\beta_c \\
      k_c^{(2\beta -\beta_c)E_0} & \beta_c/2<\beta<\beta_c \\
      \mbox{const.} & \beta<\beta_c/2 \\
    \end{array}
  \right. .
  \label{eq:escape}
\end{eqnarray}
As noted in the previous paragraph for the equilibration time,
we note that the scaling for $\beta>\beta_c$ differs from the
form $k_c^{\beta E_0}$
encountered for local trapping \cite{Baronchelli:2009}. Figure
\ref{f:escape} reports simulation data that confirm the validity of
Eq.~(\ref{eq:escape}). As the temperature is lowered, the initial
transient becomes longer, but for large enough times $t_w$ the
asymptotic behavior predicted in Eq.~(\ref{eq:escape})
is reached, as made clear from the collapse of curves concerning
different system sizes.

In the case of barrier-mediated dynamics, $\tau_k$ depends on the
symmetric function $\sigma(k,k')$, as expressed in Eq.~(\ref{eq:29}),
namely
\begin{equation}
  \label{eq:30}
  \tau_k \sim  k^{\beta E_0} e^{\beta \sigma_0 k^\mu}.
\end{equation}
Proceeding as above we obtain
\begin{eqnarray}
  t_{esc}\sim\left\{
    \begin{array}{lc}
      k_c^{(1+\epsilon)\beta E_0} & \beta>\beta_c
      \\ k_c^{[(2+\epsilon)\beta-\beta_c]E_0} & \beta_c/(2+\epsilon) <
      \beta < \beta_c \\ \mbox{const.}  & \beta<\beta_c/(2+\epsilon)
      \\
    \end{array}
  \right.  ,
\end{eqnarray}
where we recall that $\epsilon$ does not depend on the
system size.

\subsection{Average rest time}

\begin{figure}[t]
\begin{center}
\includegraphics*[width=9cm]{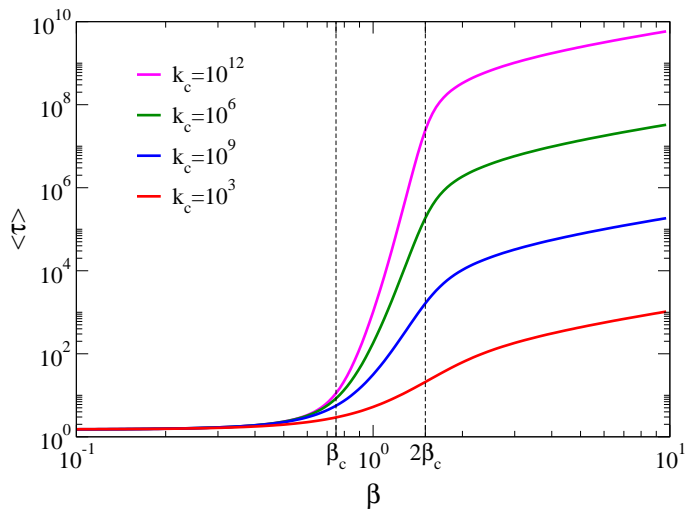}
\caption{Average rest time for the Glauber dynamics in a scale-free
  uncorrelated network with $\gamma=2.75$, as a function of the
  inverse temperature, for different system sizes. Data are obtained
  by numerical computation of Eq. (\ref{tau_zi}) (with $E_0=1$). Note
  the exponential increase with $\beta$ for $\beta_c < \beta <
  2\beta_c$, which saturates for $\beta > 2\beta_c$ as predicted by
  Eq. (\ref{e:rest}).}
\label{2temperatures}
\end{center}
\end{figure}

\begin{figure}[t]
\begin{center}
\includegraphics*[width=9cm]{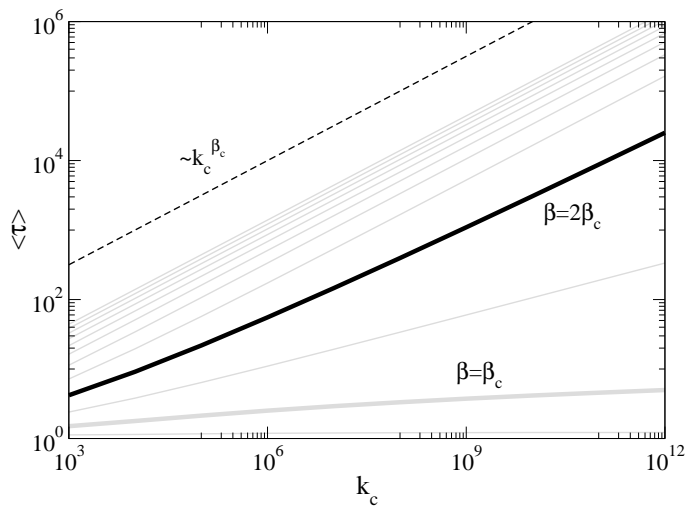}
\caption{Average rest time for the Glauber dynamics in a scale-free
  uncorrelated network with $\gamma=2.5$, as a function of the degree
  distribution cut-off and for different temperatures. Data are
  obtained by numerical computation of Eq. (\ref{tau_zi}). $\langle
  \tau \rangle$ grows as a power-law of $k_c$, with an exponent that
  grows as $\beta$ increases (going from bottom to top in the
  figure). The thick-gray line corresponds to $\beta=\beta_c$, while
  the thick-black line corresponds to $\beta=2\beta_c$. For larger
  values of $\beta$, the power-law behavior corresponds to the
  predicted $\langle \tau_k\rangle \sim k_c^{\beta_c E_0} \sim
  k_c^{\gamma-2}$, which no more depends on $\beta$. The
  $k_c^{\gamma-2}$ curve is reported as a dashed line for
  reference. Values of $\beta$ represented here are comprised between
  $0.25$ and $3$.}
\label{kcscaling}
\end{center}
\end{figure}

The HMF expression for the asymptotic average rest time, defined as
the average time spent by the system in a minimum, is given by
Eq.~(\ref{tau_zi}), namely $\langle \tau \rangle = \mathcal{Z}/
\mathcal{I}$, where the quantities $\mathcal{Z}$ and $\mathcal{I}$,
for uncorrelated scale-free networks and a degree-energy relation
$h(k)=E_0 \ln(k)$, take the form, in the continuous degree
approximation,
\begin{eqnarray}
  \label{eq:26}
  \mathcal{Z} &\sim & \int^{k_c} k^{1-\gamma +\beta E_0} \;dk \sim
  \mbox{const} + k_c^{(\beta-\beta_c)E_0} , \\
  \label{eq:27}
  \mathcal{I} &\sim& \int^{k_c} \;dk \int^{k_c}\; dk' \; k^{1-\gamma}
  g(k) k'^{1-\gamma}g(k') s(k,k').
\end{eqnarray}

Let us first recall the case of the local trap model.  Both $g(k)$ and
$s(k,k')$ are then constants, so that $ \mathcal{I} \sim \avk^2 =
\mathrm{const}$. Thus, the average rest time behaves as $\mathcal{Z}$:
it is finite for $\beta < \beta_c$, and diverges with the system size
as $k_c^{(\beta-\beta_c)E_0}$ for $\beta > \beta_c$, signaling the
emergence of the glassy regime at low temperatures.

In the case of Glauber and Metropolis dynamics (both leading to the
same results), the situation is more involved, since the
product $g(k)g(k')s(k,k')$ entering $\mathcal{I}$ is not constant. 
In fact, $\mathcal{I}$ diverges with $k_c$ for 
$\beta E_0 > 2(\gamma-2)$, that is, at a lower
temperature given by $\beta'_c=2 \beta_c$. The interplay of these two
temperatures determines the behavior of the system for finite sizes 
within the low temperature phase.
In particular, for the Glauber dynamics
with $g(k) = e^{\beta h(k)}$ and $s(k,k') =r_0/[e^{\beta h(k)
}+e^{\beta h(k')}]$, we have
\begin{equation}
  \mathcal{I}\sim  \mbox{const}+k_c^{(\beta-2\beta_c)E_0}. 
\end{equation}
Upon considering lower values of $\beta$, $\langle \tau \rangle$ first
encounters the divergence of $\mathcal{Z}$ at $\beta_c$, which is then
partially regularized by the divergence of $\mathcal{I}$ at
$2\beta_c$. From these results, we obtain the emergence of three
scaling regimes for the behavior of $\langle \tau \rangle$ as a
function of the system size:
\begin{eqnarray}
  \langle \tau \rangle^{\infty} \equiv \langle \tau \rangle (t_w \rightarrow \infty) \sim\left\{
    \begin{array}{lc}
      k_c^{\beta_cE_0} & \beta>2 \beta_c \\ 
      k_c^{(\beta-\beta_c)E_0} & \beta_c<\beta<2 \beta_c \\
      \mbox{const.} & \beta<\beta_c \\
    \end{array}\right. . 
    \label{e:rest}
\end{eqnarray}

\begin{figure}[thb]
\begin{center}
\includegraphics*[width=9cm]{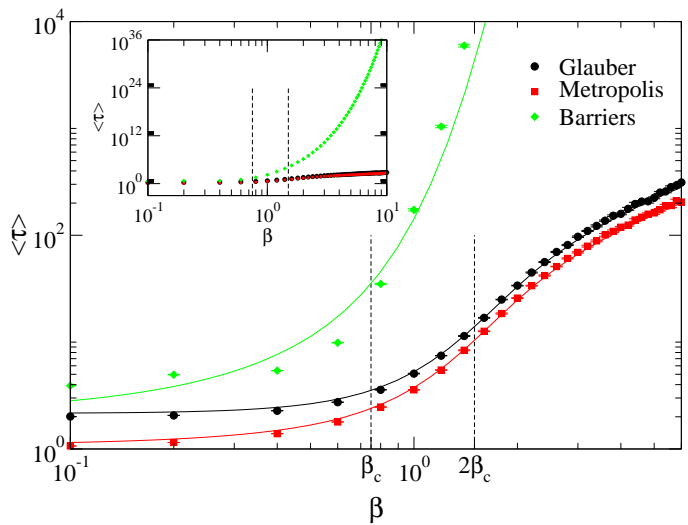}
\caption{Average rest time as a function of $\beta$ for different
  transition rates, for a random walker on
  UCM networks with $\gamma=2.75$ ($E_0=1$) and $N=10^6$. 
  Discrete points: simulation results; 
  continuous lines: theoretical predictions from $\langle \tau \rangle = \mathcal{Z}/
  \mathcal{I}$, 
  based on simulation parameters. 
  The Glauber and Metropolis
  transition rates induce two changes of behavior at $\beta_c$ and
  at $2\beta_c$, the first being a steep increase of the average rest
  time $\langle \tau \rangle$ and the second a smoothing/saturation of this
  increase. No saturation of 
  $\langle \tau \rangle$ is instead observed when barriers are present.
  The agreement with theoretical predictions is remarkable, thus corroborating 
  the validity of the HMF assumptions. Moderate deviations are found only in 
  the case of barriers, where exponential growth is expected to add greater 
  fluctuations. 
  In the inset, the difference between the two 
  behaviors is more evident thanks to a different scale of the
  plot. Note that, since in the simulations $E_0=1$, the high
  temperature limit of the rest time is different for the Glauber and
  Metropolis dynamics, being $\tau(\beta=0)=2$ and $\tau(\beta=0)=1$
  respectively. For the case of barriers we have chosen
  $\sigma_0=10^{-1}$. Each point is obtained by averaging the rest
  times corresponding to the first $10^6$ hops of the random walker in
  each of $10$ network realizations.  }
\label{simul_2temperatures}
\end{center}
\end{figure}
The direct numerical computation of Eq. (\ref{tau_zi}) is shown in
Figs. \ref{2temperatures} and \ref{kcscaling}, showing the validity of
this analysis. In particular, the exponential increase of $\langle
\tau \rangle$ with $\beta$ in the intermediate temperature range
$\beta_c<\beta<2 \beta_c$ is clearly apparent in
Fig. \ref{2temperatures}, and Fig. \ref{kcscaling} confirms that, for
$\beta>2 \beta_c$, the exponent in the scaling law for the system size
$k_c$ does not depend on the temperature. While the temperature
$\beta_c$ signals the onset of the low temperature phase with glassy
dynamics for all considered transition rates, 
for Glauber/Metropolis dynamics the low temperature phase can be further
divided into two regions that correspond to different behaviors of the
timescales with the system size.

\begin{figure}[thb]
  \begin{center}
    \includegraphics*[width=9cm]{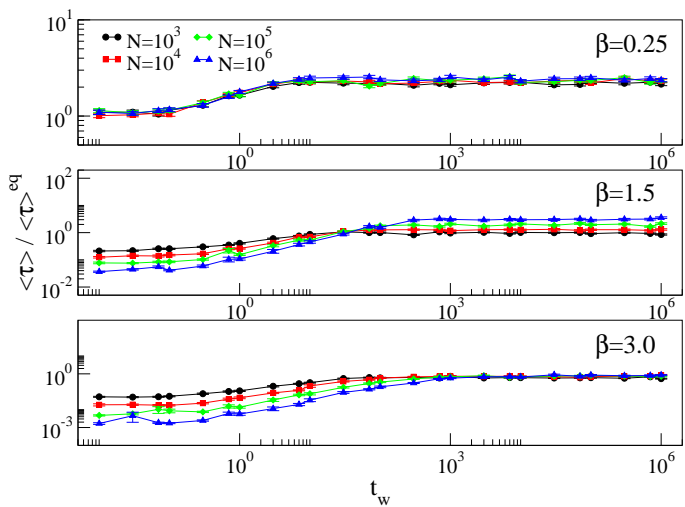}
      \caption{Average rest time for Metropolis dynamics on
        uncorrelated scale-free networks with $\gamma=3.0$ ($E_0=1$,
        $\beta_c=1.0$). Data for different system sizes collapse well
        when rescaled according to the theoretical values of
        Eq.~(\ref{e:rest}). While the agreement is excellent both for
        high and low temperature (top and bottom panels,
        respectively), logarithmic corrections are probably present
        for the regime of intermediate temperatures $\beta_c < \beta <
        2\beta_c$ (central panel). In each simulation run the rest
        interval starting before $t_w$ and ending after $t_w$ is
        considered, and each point in the Figure is averaged over
        $400$ simulation runs ($20$ runs on each of $20$ network
        instances). }
    \label{f:rest}
  \end{center}
\end{figure}

Figures \ref{simul_2temperatures} and \ref{f:rest} moreover show
the result of numerical simulations of random walkers on scale-free
networks for Glauber and Metropolis dynamics as well as in the case of
barriers, globally confirming the above discussed picture.

Dynamics in the presence of barriers do not yield the same 
phenomenology as Glauber and Metropolis rate. In this case, we have
$g(k)=1$ and $s(k,k')=r_0 e^{-\beta\sigma(k,k')}$. Selecting
$\sigma(k,k')=\sigma_0 (k^{\mu} + k'^{\mu})$, as in
Section~\ref{sec:glassy-dynamics-1}, we are led to 
\begin{equation}
  \label{eq:28}
  \mathcal{I} \sim \left[ \int^{k_c} \;dk k^{1-\gamma}  e^{-\beta
      \sigma_0 k^\mu}  \right]^2.
\end{equation}
As for the escape time $t_{esc}$, upon choosing $\sigma$ size
independent, the rest time $\langle \tau\rangle$ will be diverging
exponentially with $k_c$ at every temperature. By introducing the size
dependence as in Eq. (\ref{eq:size_depen}), instead, one can see that
the $\mathcal{I}$ integral converges to a constant for large $k_c$ so
that one is left with
\begin{eqnarray}
  \langle \tau\rangle \sim \left\{
    \begin{array}{lc}
      k_c^{(\beta-\beta_c)E_0} & \beta>\beta_c \\
      \mbox{const.} & \beta<\beta_c \\
    \end{array}
  \right. . 
\end{eqnarray}
We therefore obtain the same picture as in the case of traps, with
an exponential increase of $\langle \tau\rangle$ as $\beta$ increases,
as confirmed by numerical simulations in Fig. \ref{simul_2temperatures}.


\section{Energy basins and energy barriers}\label{sec:barriers}

Inspired by analogies with systems governed by the Arrhenius
law, we have introduced a transition rate that takes into account the
energy barriers between states.  Within the heterogeneous mean-field
approximation, in which all variables depend only on the degree of the
vertices, and choosing $E_i=E_0\ln(k_i)$, the transition rate we have
considered becomes
\begin{equation}
r(k\rightarrow k')=k^{-\beta E_0}e^{-\beta \sigma(k,k')},
\end{equation}
where $\sigma(k,k')$ is a symmetric function of the degrees
of the two nodes.  This model represents in essence an extension of
the local trap model, where non-locality enters only in the form of
symmetric energy gaps $\Sigma_{ij}$ (see Fig.~\ref{landscape}). The
steady state has exactly the same form as the ones discussed so far,
which incidentally is the same as for the local trap model. As shown
in the previous Sections, the presence of barriers
affect transient relaxation phenomena, but not the
steady state.

A different question is whether one can be more specific about the
realistic functional form of the coarse-grained function
$\sigma(k,k')$.  In the previous section we have already introduced a
definition of $\sigma(k,k')$. Here we provide the rationale behind
that choice. 

Numerical simulations of the energy-landscape network of Lennard-Jones
clusters show that the average barrier to escape from state $k$ follow
the power law $\overline{\Delta E}_k\sim k^{\mu}$, with $\mu>0$
\cite{Carmi:2009}. In our model, such average can be computed as
\begin{equation}\label{avg_barrier}
\overline{\Delta E}_k = 
\sum_h P(h|k) \left[E_0 \ln k +  \sigma(h,k)\right].
\end{equation}
For simplicity we focus on uncorrelated networks, as simulations
indeed show weak degree correlations. Under this assumption, the first
term of the sum on the right-hand side of Eq. (\ref{avg_barrier}) will
contribute as a logarithm of $k$ and the power-law behavior of
$\overline{\Delta E}_k$ is possible whenever $\sigma(h,k)\sim
k^\mu$, which leads to consider the form proposed in previous sections,
\begin{equation}\label{sigma_linear}
  \sigma(k,k')=\sigma_0 \left(k^\mu+k'^\mu\right),
\end{equation}
where $\sigma_0 $ has the dimensions of an energy (a
discussion about the possible values of $\sigma_0$ is given in Section
\ref{sec:glassy-dynamics-1}). More complicated functional forms can also be
proposed, for example accounting for barriers of different signs, as
long as they retain the same power-law behavior of
Eq. (\ref{sigma_linear}) in the large $k$ limit.  It is interesting to
notice that $\overline{\Delta E}_k\sim k^{\mu}$ implies that the
average escape rate $e^{-\beta \overline{\Delta E}_k}$ has the form of
a stretched exponential $\sim \exp(-\beta k^\mu)$, if we neglect the
logarithmic correction.

\section{Conclusions}
\label{sec:conclusions}

In this paper, we have presented a simple mathematical framework for
the description of the dynamics of glassy systems in terms of a random
walk in a complex energy landscape. We have shown how to incorporate
into this picture the network representation of this landscape, put
forward and studied by several authors
\cite{Doye:2002,Massen:2005,Seyed:2008,network_as_a_tool_1,network_as_a_tool_2,network_as_a_tool_3,network_as_a_tool_4,network_as_a_tool_5},
in order to go beyond simple mean-field models of random walks between
traps that are all connected to each other. While our previous work
had focused on the case of a landscape consisting of traps connected
by a network \cite{Baronchelli:2009}, we have here generalized our
study to more involved and realistic transition rates between minima,
including Glauber or Metropolis rates, and the possibility of energy
barriers between minima. We have shown how the interplay between the
topology of the network of minima and the relationship between the
energy and the degree of a minimum may determine a rich phenomenology,
with the existence of two phases and of glassy dynamics at low
temperature. Interestingly, the existence of these phases, and the
transition temperature, do not depend on the network's degree
correlations nor on the precise form of the transition rates, but
other more detailed properties do. In the case of Glauber and
Metropols dynamics, the low temperature phase can be further divided
into two regions with different scaling properties of the average
trapping time as a function of the temperature.  Overall, our results
rationalize and link the empirical findings about correlations between
the energy of the minima and their degree, and should stimulate
further investigations on this issue.

Our work has also interesting applications in terms of diffusion
phenomena on complex networks, and shows that non trivial transition
rates can lead to a very interesting phenomenology. Usual random walks
lead to a higher probability for the random walker to be in a large
degree node ($\propto k P(k)$), with respect to the random choice of a
node ($\propto P(k)$); here, the models we have studied can lead to various
stationary probabilities, such as for instance a uniform coverage
which does not depend anymore on the degree. Interestingly, the biased
random walks among traps that we have studied can even display a phase
transition phenomenon, as either a temperature parameter or the
network's properties are changed, with the possible presence of a
glassy phase with slow dynamics.

\section*{Acknowledgments} 
P. M., R.P.-S., and A. Baronchelli acknowledge financial support from
the Spanish MEC, under project FIS2010-21781-C02-01, and the Junta de
Andaluc\'{i}a, under project No. P09-FQM4682.. R.P.-S. acknowledges
additional support through ICREA Academia, funded by the Generalitat
de Catalunya. A. Baronchelli acknowledges support of Spanish MCI
through the Juan de la Cierva program funded by the European Social
Fund.

\providecommand{\newblock}{}


\begin{thebibliography}{10}
\expandafter\ifx\csname url\endcsname\relax
  \def\url#1{{\tt #1}}\fi
\expandafter\ifx\csname urlprefix\endcsname\relax\def\urlprefix{URL }\fi
\providecommand{\eprint}[2][]{\url{#2}}

\bibitem{Albert:2002}
Albert R and Barab\'asi A~L 2002 {\em Rev. Mod. Phys.\/} {\bf 74} 47--97

\bibitem{Dorogovtsev:2003}
Dorogovtsev S~N and Mendes J~F~F 2003 {\em Evolution of networks: From
  biological nets to the {I}nternet and {WWW}\/} (Oxford: Oxford University
  Press)

\bibitem{Newman:2003}
Newman M 2003 {\em SIAM Review\/} {\bf 45} 167--256

\bibitem{Pastor:2004}
Pastor-Satorras R and Vespignani A 2004 {\em Evolution and structure of the
  Internet: A statistical physics approach\/} (Cambridge: Cambridge University
  Press)

\bibitem{Caldarelli:2007}
Caldarelli G 2007 {\em {Scale-Free Networks: Complex Webs in Nature and
  Technology}\/} (Oxford: Oxford University Press)

\bibitem{Barrat:2008}
Barrat A, Barth\'{e}lemy M and Vespignani A 2008 {\em Dynamical Processes on
  Complex Networks\/} (Cambridge: Cambridge University Press)

\bibitem{Debenedetti_1}
Debenedetti P and Stillinger F 2001 {\em Nature\/} {\bf 210} 259

\bibitem{Debenedetti_2}
Barrat J~L, Feigelman M, Kurchan J and Dalibard J (eds) 2003 {\em Les Houches
  Session LXXVII, 1-26 July, 2002\/} Les Houches - Ecole d'Ete de Physique
  Theorique, Vol. 77 (Berlin: Springer)

\bibitem{Angelani:1998_1}
Angelani L, Parisi G, Ruocco G and Viliani G 1998 {\em Phys. Rev. Lett.\/} {\bf
  81} 4648--4651

\bibitem{Angelani:1998_2}
Berry R~S and Breitengraser-Kunz R 1995 {\em Phys. Rev. Lett.\/} {\bf 74}
  3951--3954

\bibitem{Bouchaud:1992}
Bouchaud J~P {1992} {\em {J. Physique I (France)}\/} {\bf {2}} {1705--1713}

\bibitem{Bouchaud:1995}
Bouchaud J and Dean D 1995 {\em J. Physique I (France)\/} {\bf 5} 265

\bibitem{Barrat:1995}
Barrat A and M\'ezard M {1995} {\em {J. Physique I (France)}\/} {\bf {5}}
  {941--947}

\bibitem{Monthus:1996}
Monthus C and Bouchaud J {1996} {\em {Journal of Physics A-Mathematical and
  General}\/} {\bf {29}} {3847--3869}

\bibitem{Bertin1}
Bertin E and Bouchaud J~P 2002 {\em J. Phys. A: Math. Gen.\/} {\bf 35} 3039

\bibitem{Bertin2}
Bertin E and Bouchaud J~P 2003 {\em Phys. Rev. E\/} {\bf 67} 065105(R)

\bibitem{Bertin3}
Bertin E 2003 {\em J. Phys. A: Math. Gen.\/} {\bf 36} 10683

\bibitem{Heuer_1}
B\"uchner S and Heuer A 2000 {\em Phys. Rev. Lett.\/} {\bf 84} 2168

\bibitem{Heuer_2}
de~Souza V and Wales D 2009 {\em J. Chem. Phys.\/} {\bf 130} 194508

\bibitem{Heuer_3}
Heuer A 2008 {\em J. Phys. Cond. Mat.\/} {\bf 20} 373101

\bibitem{Cieplak:1998_1}
Cieplak M, Henkel M, Karbowski J and Banavar J~R 1998 {\em Phys. Rev. Lett.\/}
  {\bf 80} 3654--3657

\bibitem{Cieplak:1998_2}
Bongini L, Casetti L, Livi R, Politi A and Torcini A 2009 {\em Phys. Rev. E\/}
  {\bf 79} 061925

\bibitem{Carmi:2009}
Carmi S, Havlin S, Song C, Wang K and Makse H~A 2009 {\em J. Phys. A\/} {\bf
  42} 105101

\bibitem{Scala:2001}
Scala A, Amaral L~A~N and Barth\'{e}lemy M 2001 {\em Europhysics Letters\/}
  {\bf 55} 594

\bibitem{Doye:2002}
Doye J~P~K 2002 {\em Phys. Rev. Lett.\/} {\bf 88} 238701

\bibitem{Massen:2005}
Massen C~P and Doye J~P~K 2005 {\em Phys. Rev. E\/} {\bf 71} 046101

\bibitem{Seyed:2008}
Seyed-allaei H, Seyed-allaei H and Ejtehadi M~R 2008 {\em Phys. Rev. E\/} {\bf
  77} 031105

\bibitem{Doye:2005}
Doye J~P~K and Massen C~P 2004 {\em J. Chem. Phys.\/} {\bf 122} 084105. 14 p

\bibitem{network_as_a_tool_1}
Gfeller D, De~Los~Rios P, Caflisch A and Rao F 2007 {\em Proceedings of the
  National Academy of Sciences\/} {\bf 104} 1817--1822

\bibitem{network_as_a_tool_2}
Gfeller D, de~Lachapelle D~M, De~Los~Rios P, Caldarelli G and Rao F 2007 {\em
  Phys. Rev. E\/} {\bf 76} 026113

\bibitem{network_as_a_tool_3}
Burda Z, Krzywicki A, Martin O~C and Tabor Z 2006 {\em Phys. Rev. E\/} {\bf 73}
  036110

\bibitem{network_as_a_tool_4}
Burda Z, Krzywicki A and Martin O~C 2007 {\em Phys. Rev. E\/} {\bf 76} 051107

\bibitem{network_as_a_tool_5}
Baiesi M, Bongini L, Casetti L and Tattini L 2009 {\em Phys. Rev. E\/} {\bf 80}
  011905

\bibitem{Baronchelli:2009}
Baronchelli A, Barrat A and Pastor-Satorras R 2009 {\em Phys. Rev. E\/} {\bf
  80} 020102

\bibitem{DorogoRev}
Dorogovtsev S~N, Goltsev A~V and Mendes J~F~F 2008 {\em Rev. Mod. Phys.\/} {\bf
  80} 1275--1335

\bibitem{Noh:2003}
Noh J~D and Rieger H 2004 {\em Phys. Rev. Lett.\/} {\bf 92} 118701

\bibitem{rw_weight}
Baronchelli A and Pastor-Satorras R 2010 {\em Phys. Rev. E\/} {\bf 82} 011111

\bibitem{Bortz:1975}
Bortz A, Kalos M and Lebowitz J 1975 {\em J. Comp. Phys.\/} {\bf 17} 10

\bibitem{Krauth}
Krauth W 2006 {\em Statistical Mechanics: Algorithms and Computations\/}
  (Oxford: Oxford University Press)

\bibitem{Catanzaro:2005}
Catanzaro M, {Bogu\~{n}\'{a}} M and Pastor-Satorras R 2005 {\em Phys. Rev. E\/}
  {\bf 71} 027103

\bibitem{baronchelli2008random}
Baronchelli A, Catanzaro M and Pastor-Satorras R 2008 {\em Phys. Rev. E\/} {\bf
  78} 011114

\bibitem{molloy95}
Molloy M and Reed B 1995 {\em Random Struct. Algorithms\/} {\bf 6} 161

\bibitem{alexei}
Pastor-Satorras R, V{\'a}zquez A and Vespignani A 2001 {\em Phys. Rev. Lett.\/}
  {\bf 87} 258701

\bibitem{AGA-05}
Agaev R and Chebotarev P 2005 {\em Linear Algebra and its Applications\/} {\bf
  399} 157--168

\bibitem{meyer}
Meyer C~B 2000 {\em Matrix Analysis and Applied Linear Algebra\/} (Philadephia:
  Society for Industrial and Applied Mathematics)

\bibitem{marian1}
{Bogu\~{n}\'{a}} M and Pastor-Satorras R 2002 {\em Phys. Rev. E\/} {\bf 66}
  047104

\bibitem{Barthelemy:2005}
Barth\'elemy M, Barrat A, Pastor-Satorras R and Vespignani A 2005 {\em J.
  Theor. Biol.\/} {\bf 235} 275--288

\bibitem{abramovitz}
Abramowitz M and Stegun I 1964 {\em Handbook of Mathematical Functions\/} 5th
  ed (New York: Dover)

\end{thebibliography}
\end{document}